\documentclass[12pt]{article}
\usepackage[top=1in, bottom=1in, left=1.in, right=1.in]{geometry}
\usepackage[english]{babel}
\usepackage{atbegshi,cite}
\usepackage{amsmath,amssymb,amsbsy,amstext, amsthm, simplewick}
\usepackage{hyperref}
\usepackage{graphicx}
\usepackage{amsfonts}
\usepackage[small]{caption}
\usepackage{upgreek}
\usepackage[titletoc]{appendix}
\usepackage{setspace}
\usepackage[dvipsnames]{xcolor}
\usepackage{slashed}
\usepackage{comment}
\usepackage{soul}

\newcommand{\newc}{\newcommand}

\newc{\sflip}{{\bf \color{red} ***sign***}}
\newc{\TR}{{\rm Tr}}
\newc{\Zbb}{{\mathbb Z}}
\newc{\Rt}{{\mathbb R}^3}
\newc{\Rtw}{{\mathbb R}^2}
\newc{\Rf}{{\mathbb R}^4}
\newc{\So}{{\mathbb S}^1}
\newc{\Stw}{{\mathbb S}^2}
\newc{\Sth}{{\mathbb S}^3}
\newc{\zt}{{\mathbb Z}_2}
\newc{\RtSo}{{\mathbb R}^3\times{\mathbb S}^1}
\newc{\RtwStw}{{\mathbb R}^2\times{\mathbb S}^2}
\newc{\scriminus}{{\cal I}^-}
\newc{\scriplus}{{\cal I}^+}
\newc{\mpl}{M_p}
\newc{\Ricci}{\mathcal{R}}
\newc{\calU}{{\cal U}}
\newc{\calK}{K}
\newc{\calUi}{{\cal U}^{-1}}
\newc{\calG}{{\cal G}}
\newc{\calM}{{\cal M}}
\newc{\calT}{{\cal T}}
\newc{\calL}{{\cal L}}
\newc{\calO}{{\cal O}}
\newc{\calR}{{\cal R}}
\newc{\calH}{{\cal H}}
\newc{\calQ}{{\cal Q}}
\newc{\calI}{{\cal I}}
\newc{\calOb}{{\cal O}^\dagger}
\newc{\IEH}{I_{\calL}}
\newc{\Irad}{I_{\rm ADM}^{\rm radial}}
\newc{\Itemp}{I_{\rm ADM}^{\rm temporal}}

\begin{document}
\begin{titlepage}
\begin{flushright}
{\large 
~\\
}
\end{flushright}

\vskip 2.2cm

\begin{center}

{\large \bf de Sitter Black Holes as Constrained States\\
in the Euclidean Path Integral}

\vskip 1.4cm

{ Patrick Draper$^{(a)}$ and Szilard Farkas}
\\
\vskip 1cm
{$^{(a)}$ Department of Physics, University of Illinois, Urbana, IL 61801}\\
\vspace{0.3cm}
\vskip 4pt

\vskip 1.5cm

\begin{abstract}
Schwarzschild-de Sitter black holes have two horizons that are  at different temperatures for generic values of the black hole mass. Since the horizons are out of equilibrium the solutions do not admit a smooth Euclidean continuation and it is not immediately clear what role they play in the gravitational path integral. We show that Euclidean SdS is a genuine saddle point of a certain constrained path integral, providing
 a consistent Euclidean computation of the probability  $\sim e^{-(S_{dS}-S_{SdS})}$ to find a black hole in the de Sitter bath.

\end{abstract}

\end{center}

\vskip 1.0 cm

\end{titlepage}
\setcounter{footnote}{0} 
\setcounter{page}{1}
\setcounter{section}{0} \setcounter{subsection}{0}
\setcounter{subsubsection}{0}
\setcounter{figure}{0}

\section{Introduction}
A surprising property of de Sitter space is that  local excitation lower the total entropy. In the case of black holes, the sum of the cosmological and black hole horizon areas in Schwarzschild-de Sitter (SdS) solutions  decreases as the black hole mass increases. On these grounds it has been suggested that local excitations should be thought of as constrained states of horizon degrees of freedom in empty de Sitter~\cite{Banks:2006rx,Banks:2013fr} (see also more recently~\cite{Johnson:2019ayc,Johnson:2019vqf,Dinsmore:2019elr,Banks:2020zcr,Susskind:2021dfc}. Here we show that the constrained state idea is also useful to describe the contribution of black holes to the bulk Euclidean path integral.

In static coordinates the Euclidean continuation of SdS is
\begin{align}
ds^2&= f(\rho) dt^2 + f(\rho)^{-1} d\rho^2+\rho^2 d\Omega^2\nonumber\\
f(\rho)&=1-2M/\rho-(\rho/L)^2.
\label{eq:sdsmetric}
\end{align}
$L$ is the de Sitter radius and $M$ is the black hole mass parameter. The black hole and cosmological horizons $r_{b,c}$ are located at the zeroes of $f$, and this metric has an irremovable conical singularity at one of them. More precisely, each horizon has a potential conical singularity, and typically only one of the two can be removed by a choice of periodicity of $t$. For this reason Euclidean SdS does not provide a useful saddle point of the ordinary path integral, except in the Nariai limit, where the temperatures coincide and the solution is smooth~\cite{Ginsparg:1982rs}. 

However, Euclidean SdS does provide a genuine saddle point of a constrained path integral. The constraint is the specification of boundary data on a sphere that uniquely determines the mass of the enclosed black hole. This constraint can easily be imposed if the path integral is decomposed into path integrals over fields in the two regions separated by the sphere, similarly to the factorization of transition amplitudes into regions separated by a time slice. In this note we study this factorization and use it to show that black holes contribute to the de Sitter partition function with the  probability weight $e^{S_{SdS}-S_{dS}}$. This result can be anticipated on other grounds, but our goal is to explain how it arises from the semiclassical approximation to the gravitational path integral.

This note is a companion to~\cite{DFeuclidean1} in which we discuss some other properties of Euclidean SdS in four dimensions, focusing on cavity partition functions and equilibrium thermodynamics. Recent interesting work related to this includes~\cite{svesko2022quasilocal}, which considers similar physics in two dimensions and in finite causal diamonds~\cite{JV,Banks:2020tox}, and~\cite{Marolf:2022ntb,Marolf:2022jra} which addresses the stability problem for cavity black holes with zero or negative cosmological constant.

We would also like to draw attention to upcoming work by Visser, van der Schaar, and Morvan on Euclidean SdS and constrained path integrals, very similar  to the ideas considered here, which we learned of after this paper was completed. We thank them for letting us know of their work.

\section{Factorization and boundary terms in the path integral}
Before examining SdS specifically, let us take a step back and review some properties of the gravitational path integral, particularly the factorization properties with fixed boundary data in the Hamiltonian formalism. In writing the path integral, we can use either the Euclidean Lagrangian action $I_{\calL}$, containing both the Einstein-Hilbert (EH) action and the Gibbons-Hawking-York boundary term, or the Euclidean Arnowitt-Deser-Misner (ADM) action with certain boundary terms $B_N$ and $B_{N_a}$:
\begin{align}
\IEH &= -\frac{1}{16\pi}\int_{\cal{M}}d^4x\sqrt{g}(R-2\Lambda)-\frac{1}{8\pi}\int_{\partial\cal M}d^3x\sqrt{\gamma}K,\label{eq:IE}\\
I_{\rm ADM}&=\frac{1}{16\pi }\int_{t_1}^{t_2} dt\int_{\Sigma_t}d^3x\,\left(\pi^{ab}\partial_t h_{ab}-N{\cal H}-N_a{\cal H}^a\right)+B_N+B_{N_a}.\label{eq:IADM}
\end{align}
For our purposes, it will be sufficient to focus on manifolds of fixed topology. The path integral in the Hamiltonian formalism is
\begin{align}
Z=\int Dh_{ab}\,D\pi^{ab} D\!N D\!N_a\, e^{-I_{\rm ADM}[h_{ab},\pi^{ab},N,N_a]}.
\end{align}
The Hamiltonian and momentum constraints are\footnote{Note that the relative sign of the curvature and momentum terms differs between Euclidean and Lorentzian signature.}
\begin{align}
\calH = {}^{(3)\!}R + \frac{1}{h}&\left(\pi^{ab}\pi_{ab}-\frac{1}{2}\pi^2\right),\\
\calH^b = &-2D_a\frac{\pi^{ab}}{\sqrt{h}},
\end{align}
where $D_a$ is the covariant derivative compatible with the induced metric $h_{ab}$  on $\Sigma_t$.

We will be interested  in the path integral over metrics on $\calM=S^4$, for which the saddle point is empty Euclidean de Sitter and we denote the partition function by $Z_{\rm dS}$, as well as in path integrals on subregions of $S^4$. In the former case there are no boundaries and so boundary terms play no role, but in the latter case there will be boundaries that partition the manifold. The Euclidean time foliation is also periodic, and therefore we omit the limits on the $t$ integrals going forward.

Different choices for the boundary terms $B_N$ and $B_{N_a}$ correspond not only to different classical variational problems, but also to different ways in which the full path integral can be split into path integrals over fields in different parts of $\calM$. We discuss first the standard choice of boundary terms, related to the canonical ensemble, and then an alternate choice, related to the microcanonical ensemble.

\subsection{Canonical boundary data}
The boundary terms $B_N$ and $B_{N_a}$ are usually chosen so that $I_{\rm ADM}$ has a well-posed variational problem with fixed $h_{ab}$, $N$, and $N_a$ at the boundary. They can be derived by requiring that their variations cancel the surface terms generated by the variations of the bulk term. Let us introduce a radial coordinate $\rho$, analogous to the one that appears in the static coordinate representation of de Sitter, although we make no assumptions about metric symmetries at this stage.  
Using the Gauss-Codazzi relations, we can write the spatial curvature scalar ${}^{(3)\!}R$  as
\begin{align}
{}^{(3)\!}R={}^{(2)\!}R+k^2-k_{ab}k^{ab}-2D_a(r^ak)+2D_a(r^bD_br^a),\label{eq:3R_decomp}    
\end{align}
where $r^a$ is the unit normal to surfaces of fixed $\rho$,  $s_{ab}$ and ${}^{(2)}R$ are the induced metric and curvature scalar, $k_{ab}=s_a^cD_cr_b$ is the extrinsic curvature, and $k=k_a^a$. Let $\cal{T}$ be a surface defined by $\rho=r$ and consider the action on one side of $\cal{T}$. Using Eq.~\eqref{eq:3R_decomp}, the integral of $N{\cal H}$ can be written as the bulk integral of an expression containing at most first order radial derivatives of $N$ and $h_{ab}$, plus a boundary term
\begin{align}
2\int dt\int_{S_t}d^2x\sqrt{s}(Nk-r_ar^bD_br^a),
\label{eq:NHibp}
\end{align}
where $S_t$ is the intersection of $\cal{T}$ and $\Sigma_t$. Since $r^ar_a=1$, the second term in the integrand vanishes. So we can cancel the contribution of~\eqref{eq:NHibp} to the action by the choice
\begin{align}
B_N=-\frac{1}{8\pi}\int dt\int_{S_t}d^2x\sqrt{s}Nk~~~~~~~{\rm (canonical)}\label{eq:BN} 
\end{align}
Similarly, the sum of the momentum constraint term in $I_{\rm ADM}$ and the boundary term
\begin{align}
B_{N_a}=-\frac{1}{8\pi}\int dt\int_{S_t}d^2x\sqrt{s}N_a\frac{\pi^{ab}}{\sqrt{h}}r_b\label{eq:BNa}  
\end{align}
is proportional to  the bulk integral of $\sqrt{s}\pi^{ab}D_aN_b$.
Therefore, with the standard boundary terms \eqref{eq:BN} and \eqref{eq:BNa}, the variational problem of $I_{\rm ADM}$ is well-posed with fixed $h_{ab}$, $N$, and $N_a$ on $\cal T$.  It follows that these boundary terms allow us to write $I_{\rm ADM}$ as the bulk integral of a function of the field variables and their derivatives, with at most first order radial derivatives.

Now let us consider both sides of $\calT$. $\calT$ divides  $\calM$ into two regions, $\calM_1$ and $\calM_2$. 
We would like to write the path integral over fields defined on $\calM$ as a product of path integrals over fields on $\calM_1$ and $\calM_2$, with a lower dimensional path integral over data at the interface. The result, given below in Eq.~\eqref{eq:fact_canonical}, is in some sense what we might expect, but in fact it is intricately tied to the choice of the boundary terms. Since we will change the boundary terms in the next section, we start by elaborating on this connection.

One way to approach the problem of factorization is to consider the domain of the action functional appearing in the path integral. Of course, the action is well-defined -- i.e. without specifying additional UV data -- for smooth metrics, but it is also well-defined in this sense for a larger space of field variables. We have some freedom in this extension. 

We focus on the continuity properties of the fields at $\calT$. If we take $s_{ab}$, $N$, and $N_a$ on fixed $\rho$ surfaces to be continuous (in $\rho$) at $\calT$, then it is not necessary  for $k$ and $\pi^{ab}$  to be continuous (in $\rho$) at $\calT$ in order for the ADM action 
to be well-defined. There may be Dirac delta functions in the bulk integrand, but they are multiplied by continuous functions of $\rho$, and so are integrable without regularization. 

Furthermore, we have seen that the action can be written as a bulk integral with at most first order radial derivatives of the field variables (in fact no derivatives of $\pi^{ab}$). In this form, there will be no Dirac delta functions in the action density in $\calM$ even if $k$ and $\pi^{ab}$ are discontinuous, merely products of discontinuous functions. 
 The action on all of $\calM$ can be written in this way, as can the action on each of the subregions $\calM_1$ and $\calM_2$. Therefore, in this form, the action functional manifestly satisfies the additivity property
\begin{align}
I_{\rm ADM}(\calM) = I_{\rm ADM}(\calM_1) + I_{\rm ADM}(\calM_2).\label{eq:additive}
\end{align}
This decomposition is immediate when using the ``purely bulk" form of the action in the subregions, and therefore it also holds in the equivalent bulk+boundary form, for boundaries placed just on either side of $\calT$. 

Let us illustrate these comments more explicitly with the example of discontinuous $k$. Using Eq.~\eqref{eq:3R_decomp}, the ADM action of a metric on all of $\calM$ contains terms that can be expressed in different ways:
\begin{align}
I_{\rm ADM}&\supset+ \frac{1}{8\pi}\int_{\calM}\, N D_a(r^a k)\nonumber\\
&= -\frac{1}{8\pi}\int_{\calM}\, (D_aN) r^a k\nonumber\\
&= -\frac{1}{8\pi}\int_{\calM_1}\, (D_aN) r^a k-\frac{1}{8\pi}\int_{\calM_2}\, (D_aN) r^a k\nonumber\\
&= +\frac{1}{8\pi}\int_{\calM_1}\, N D_a(r^a k)+ \frac{1}{8\pi}\int_{\calM_2}\, N D_a(r^a k)-\frac{1}{8\pi}\int_{\calT^-}d^3x\sqrt{s}Nk-\frac{1}{8\pi}\int_{\calT^+}d^3x\sqrt{s}Nk.
\end{align}
There are no boundary terms in the first line because $\calM$ does not have a boundary. In this line the  integrand  contains radial derivatives of $k$, and so exhibits a delta function singularity if $k$ has a discontinuity in $\rho$ at $\calT$. However, it is multiplied by $N$, which is continuous at the singularity and so the integral remains well-defined. In the second line,  both $\partial_\rho N$ and $k$ can be discontinuous in $\rho$ at $\calT$, but again the action functional is well-defined, and moreover there is no delta function singularity. This makes the split into subregions trivial in the third line. In the fourth line we return to the original form of the action in the subregions, which has boundary terms on both sides of $\calT$, denoted $\calT^\pm$. The boundaries excise the delta function singularity and its contribution is reproduced by the boundary terms.

With Eq.~\eqref{eq:additive}, it is clear how the path integral decomposes. Variables  that are required to be continuous in $\rho$ at $\calT$ are shared between the two regions, and therefore must be integrated over only once. We can write the factorization as:
\begin{align}
Z_{\rm dS}=\int_\calT Ds_{ab}\,D\!N D\!N_a\,Z(\calM_1|s_{ab},N,N_a)Z(s_{ab},N,N_a|\calM_2).\label{eq:fact_canonical}
\end{align}
where $\calT$ in the subscript of the integral  is a mnemonic that integration variables are  defined on the boundary $\calT$. We indicated explicitly the dependence of the path integrals in $\calM_1$ and $\calM_2$ on the boundary data at $\calT$, using notation suggestive of the insertion of a complete set of states in quantum mechanics. Since $Z(\calM_1)$ and $Z(\calM_2)$ are otherwise independent, it is clear that $Z_{\rm dS}$ receives contributions from metrics on $\calM$ in which some properties can be discontinuous -- for example, $h_{\rho\rho}$ may be discontinuous in $\rho$ at $\calT$, leading to discontinuous $k$, as discussed above. Below, we will consider a different but analogous factorization. 

The boundary data over which we integrate in Eq.~\eqref{eq:fact_canonical} includes only the metric $s_{ab}$ induced on $S_t=\Sigma_t\cap\calT$, even though the above argument suggests that the entire spatial metric $h_{ab}$ needs to be continuous at $\calT$ for Eq.~\eqref{eq:additive} to hold. 
In the Hamiltonian formalism, it is convenient to impose the gauge condition 
\begin{align}
h_{\rho\alpha}=0\;\mbox{for}\;\alpha\neq\rho\;\mbox{at}\;\calT,\label{eq:gauge_gra}
\end{align}
and then there is no integration with respect to the boundary value of these components. This condition is not restrictive: since it is imposed only on the boundary, the stationarity of the action still provides the full set of field equations in the bulk. Furthermore, for any solution, the gauge condition \eqref{eq:gauge_gra} can always be satisfied by an appropriate foliation in the neighborhood of $\calT$. We will assume that this gauge condition is satisfied. It also ensures that $I_{\rm ADM}$ with the boundary terms \eqref{eq:BN} and \eqref{eq:BNa} is equal to the Euclidean Lagrangian action $I_\calL$. 
The variational problem of $I_\calL$ is well posed even if only the induced metric on the boundary is fixed, so the components that are not part of the induced metric must be undifferentiated with respect to the coordinate off the boundary. Put more simply, there are no $\partial_\rho h_{\rho\rho}(\calT)$ terms in the action.  Correspondingly, in the Lagrangian formalism, the factorization property analogous to Eq.~\eqref{eq:fact_canonical} has an integral over the induced metric on $\calT$, encoded by the field variables ($s_{ab}$, $N$, $N_a$) in the Hamiltonian formalism. The path integral over the other components of the metric can be performed independently in the two regions and therefore is included in the definition of $Z(\calM_1)$ and $Z(\calM_2)$.

\subsection{Microcanonical boundary data}

The factorization in \eqref{eq:fact_canonical} is not the only interesting case. Now we discuss a different decomposition which is connected to a different set of boundary terms. First, note that Eq.~\eqref{eq:3R_decomp} contains only first order radial derivatives of the metric, apart from the term proportional to $D_a(r^a k)$.\footnote{
The radial derivatives in the  last term in Eq.~\eqref{eq:3R_decomp} are only first order because $r^ar_a=1$ implies that $A^a=r^bD_br^a$ is orthogonal to $r^a$, so its divergence $D_aA^a$ contains derivatives of $A^a$ only in directions tangent to the constant $\rho$ surfaces.}
Instead of the boundary term in \eqref{eq:BN}, let us take
\begin{align}
B_N=0~~~~~~~{\rm (microcanonical)}\label{eq:BN0}
\end{align}
 With this choice,  the term $-2N D_a(r^ak)$ in the action density is no longer effectively integrated by parts.  We still choose $B_{N_a}$ as in Eq.~\eqref{eq:BNa} (this is why ``canonical" appears in Eq.~\eqref{eq:BN} but not in Eq.~\eqref{eq:BNa}), and
 then we only need to integrate by parts in the momentum constraint term to cancel out $B_{N_a}$ and write $I_{\rm ADM}$ as a purely bulk integral.

What are the factorization properties of the path integral with these boundary terms? Again let us appeal to the action functional. Take $s_{ab}$, $k$, and $N_a$ to be continuous in $\rho$ at $\calT$. (In the gauge~\eqref{eq:gauge_gra}, continuity of $h_{ab}$ amounts to continuity of $s_{ab}$.) Then the action is well-defined without additional UV data even if $N$ is discontinuous.
The  additivity property \eqref{eq:additive} is manifest already in the ordinary form of the ADM action:
\begin{align}
I_{\rm ADM}&\supset \frac{1}{8\pi}\int_{\calM}\, N D_a(r^a k)\nonumber\\
&= \frac{1}{8\pi}\int_{\calM_1}\, N D_a(r^a k)+ \frac{1}{8\pi}\int_{\calM_2}\, N D_a(r^a k),
\end{align}
and it implies the following factorization of the path integral:
\begin{align}
Z_{\rm dS}=\int_\calT Ds_{ab}\,Dk\, D\!N_a\,Z(\calM_1|s_{ab},k,N_a)Z(s_{ab},k,N_a|\calM_2).\label{eq:fact_micro}
\end{align}
This holds already in the ordinary form of the ADM action, Eq.~\eqref{eq:IADM}, with the new choice of boundary term $B_N=0$ on either side of $\calT$. (The measure in \eqref{eq:fact_micro} is somewhat schematic, but it is not important for our purposes.)

Let us pause and make two comments:

\begin{itemize}
\item Eq.~\eqref{eq:fact_micro} and \eqref{eq:fact_canonical} may be thought of as giving two different ``complete set of states" insertions into the same partition function $Z_{\rm dS}$.
    \item The temporal foliation of $\calT$ need not have a continuation into the bulk of both $\calM_1$ and $\calM_2$ that is nonsingular everywhere. That will be the case for Euclidean SdS and we will deal with this mild complication below. Although we used the Hamiltonian formalism in the derivation above, 
once a foliation is defined on $\calT$, we can specify the boundary data with which $Z(\calM_1|s_{ab},k,N_a)$ and $Z(s_{ab},k,N_a|\calM_2)$ are well-defined. Then they can be computed
using the Lagrangian formulation with action $I_\calL$, for which no bulk foliation is required, plus appropriate boundary terms. 

\end{itemize}

\section{SdS as a constrained state}
Now let us discuss constrained states. 
The restriction to a black hole of mass $M$ is obtained formally by inserting a  delta functional at $\calT$, concentrated on the boundary data ($s_{ab}^M$, $k^M$, $N_a^M$) of the black hole. This gives:
\begin{align}
Z_{\rm SdS}^M=Z(\calM_1|s_{ab}^M,k^M,N_a^M)Z(s_{ab}^M,k^M,N_a^M|\calM_2).\label{eq:ZSdS}
\end{align}

Before we proceed let us elaborate on the motivation for fixing $k$ instead of $N$. A priori, we could do either, and no particular justification is required for choosing one or the other. However, there is a reason to preferentially fix $k$. With the canonical factorization and a shared fixed boundary metric, we obtain two products of canonical partition functions. However, the canonical ensemble is not well-defined for spherical cavities at positive cosmological constant. The lowest-action solution in each region consistent with the boundary conditions contains a cosmological horizon and possesses a negative heat capacity~\cite{DFeuclidean1}. In contrast, the cavity solutions with fixed $k$ have a consistent microcanonical thermodynamic interpretation~\cite{DFeuclidean1}. This is one reason we proceed with the microcanonical factorization.

As in the more general discussion of Sec. 2, we adopt a  coordinate system where  $\calT$ lies at a fixed radial coordinate $\rho=r$.  On $\calT$ we set $N_a=N_a^M=0$, we fix $s_{ab}=s_{ab}^M$ to the ordinary round metric $r^2 d\Omega_2^2$, and we take $k = k^M= \frac{2}{r}\sqrt{f(r)}$ with $f$ given in Eq.~(\ref{eq:sdsmetric}).

The semiclassical approximation can now be applied to each region separately.  
Euclidean SdS is a solution in each region, and now {\emph{both}} conical singularities can be removed by an appropriate rescaling of the lapse function. Let the periodicity of the time coordinate be such that the metric~(\ref{eq:sdsmetric}) is smooth for $\rho<r$; i.e., $t\sim t+1/T_b$, where the black hole temperature $T_b(M)$ is determined by the mass $M$ consistent with the fixed value of $k$. The corresponding cosmological horizon temperature $T_c$ is likewise determined by $M$. Then the smooth semiclassical solution for $\rho>r$ and the same temporal periodicity is
\begin{align}
ds^2&= f(\rho)\left(\frac{T_b}{T_c}\right)^2 dt^2 + f(\rho)^{-1} d\rho^2+\rho^2 d\Omega^2.
\label{eq:sdsmetricout}
\end{align}
There is no conical singularity at the cosmological horizon $\rho=r_c$.  
The lapse is not continuous at $\calT$, but it does not need to be.

The momenta vanish in SdS and so one might conclude that the semiclassical action~(\ref{eq:IADM}) vanishes. This is not quite correct for reasons described in~\cite{Banados:1993qp,Teitelboim:2001skl}. As alluded to in the bullet points above, the problem is the time foliation used in the ADM action above cannot be extended everywhere into the bulk; the foliation breaks down at the two horizons. The SdS solutions are smooth at these points, but the ADM action cannot be evaluated. Fortunately, there is a simple fix~\cite{Banados:1993qp,Teitelboim:2001skl}: we just use a form of the action that is well-defined everywhere. Any such form is equally valid since this just a trick to compute the action of a solution we have already obtained. One simple prescription is as follows. 
We draw  infinitesimal boundaries $\calT_b$ and $\calT_c$ around the black hole and cosmological horizons. Between $\calT_b$ and $\calT_c$ we use the ADM form, which can be extended this far into the bulk. Then 
inside the boundaries  we use the ordinary EH+GHY action, adding also the canonical  boundary terms on the ADM side.

In the limit that the boundaries $\calT_{b,c}$ are taken infinitesimally close to the horizons, all of the bulk contributions to the SdS action vanish. The bulk ADM contributions vanish because $\pi^{ab}=0$, and the EH contributions vanish because the curvature is finite and integrated over a region of zero volume in the limit. The $B_{N_a}$ boundary terms near the horizons are also zero on SdS, and the canonical Hamiltonian boundary terms at each horizon vanish because $Nk\rightarrow 0$ at those points. Thus the entirety of the classical action is due to  GHY terms on the infinitesimal boundaries $\calT_{b,c}$: 
\begin{align}
I_{\rm \tiny tot}&=I_{\rm \tiny  GHY}(\calT_b)+I_{\rm \tiny GHY }(\calT_c)\nonumber\\
&=-(A_b+A_c)/4.
\label{eq:finalaction}
\end{align}
Happily, the result does not depend on the radius $r$ where we placed the constraint: the total action is just minus the total entropy. Regardless of where the constraint is placed, the boundary data corresponds to the same physical state in Lorentzian signature -- the unique Lorentzian SdS solution with that value of $M$ -- and so the $r$-independence of Eq.~\eqref{eq:finalaction} means that $Z_{\rm SdS}^M$ is a well-defined property of this state. More generally, it is not necessary to construct it with a spherically symmetric constraint. $\calT$ can be any static surface enclosing the black hole, and for suitable boundary data we will still obtain the same $Z_{\rm SdS}^M$. 

As a formal prescription for performing a continuation of the full discontinuous Euclidean solution to Lorentzian signature, we can proceed as follows. First we change the $t$ coordinate in one region so that the lapse is continuous but the coordinate periodicity is not. Then in each region we perform the naive continuation, which erases the discontinuity in the periodicity by decompactifying $t$ in both regions. The result is the unique SdS solution of the given mass which is smooth apart from the black hole singularity.

Thus the saddle point of the constrained path integral computes the probability to find a black hole of mass $M$ in the de Sitter ensemble,
\begin{align}
    P= Z_{\rm SdS}^M/Z_{\rm dS} \sim e^{S_b+S_c-\pi L^2},
\end{align}
normalizing to the unconstrained partition function. 
This interpretation offers a resolution to the semiclassical meaning of Euclidean SdS. The usual conical singularity can be exchanged for a discontinuity in the local temperature at some radius, encoding the absence of thermal equilibrium. The discontinuous geometry is a stationary point of the action with the constraint, and it yields the expected result that fluctuating a black hole into existence is  exponentially rare in the entropy deficit.

\section*{Acknowledgments}
This work was written with support from the US Department of Energy under grant number DE-SC0015655, and from the DOE Office of High Energy
Physics QuantISED program under an award for the Fermilab Theory Consortium “Intersections of QIS and
Theoretical Particle Physics.”

\bibliography{euclideands_refs}{}
\bibliographystyle{utphys}

\end{document}